\documentclass[prl,twocolumn,showpacs,amsmath,amssymb]{revtex4}
\input epsf

\usepackage{graphicx}

\newcommand{\cO}{{\cal O}}

\def\be{\begin{equation}}
\def\ee{\end{equation}}
\def\bea{\begin{eqnarray}}
\def\eea{\end{eqnarray}}

\begin{document}
\title{Solvable Relativistic Hydrogenlike System in Supersymmetric Yang-Mills Theory}

\author{Simon Caron-Huot}
\affiliation{Niels Bohr International Academy and Discovery Center, Blegdamsvej 17, Copenhagen 2100,
Denmark\\
and Institute for Advanced Study, Princeton, New Jersey 08540, USA}
\email{schuot@nbi.dk}
\author{Johannes M.\ Henn}
\affiliation{Institute for Advanced Study, Princeton, New Jersey 08540, USA}
\email{jmhenn@ias.edu}

\date{\today}

\begin{abstract}
The classical Kepler problem, as well as its quantum mechanical version, the hydrogen atom, enjoys a well-known hidden symmetry, the conservation of the Laplace-Runge-Lenz vector, which makes these problems superintegrable. Is there a relativistic quantum field theory extension that preserves this symmetry? In this Letter we show that the answer is positive: in the nonrelativistic limit, we identify the dual conformal symmetry of planar $\mathcal{N}=4$ super Yang-Mills theory with the well-known symmetries of the hydrogen atom.
We point out that the dual conformal symmetry offers a novel way to compute the spectrum of bound states of massive $W$ bosons in the theory. 
We perform nontrivial tests of this setup at weak and strong coupling and comment on the possible extension to arbitrary values of the coupling.
\end{abstract}

\maketitle



The classical two-body (or Kepler) problem, with Hamiltonian 
\be
H=\frac{p^2}{2\mu} -\frac{\lambda}{4 \pi} \frac{1}{|x|}\,, \label{Hamiltonian}
\ee
is well known to possess a nonobvious conserved vector which makes it {super}integrable.
This Laplace-Runge-Lenz vector is expressed as
\be
 \vec{A} = \frac12\left(\vec{p}\times \vec{L} - \vec{L}\times  \vec{p}\right)
 - \mu  \frac{\lambda}{4 \pi}   \frac{\vec x}{|x|}\,, \label{Laplace-Runge-Lenz}
\ee
where $\vec{L} = \vec{x} \times \vec{p}$ is the angular momentum.
Physically, its conservation accounts for the fact that the orbits of the $1/|x|$ central potential form closed ellipses which do not precess with time.

The same Hamiltonian is relevant for the quantum mechanical description of the hydrogen atom,
with $\vec{x}$ and $\vec{p}$ replaced by operators.
As was pointed out early on by Pauli, the Laplace-Runge-Lenz vector in the above form
is also conserved quantum mechanically; i.e., it commutes with the Hamiltonian.
The symmetry group is enlarged from $SO(3)$ rotations to $SO(4)$.
This gives rise to a simple algebraic way of calculating the spectrum, 
which automatically accounts for its degeneracies~\cite{Pauli}.

In real hydrogen atoms, both this symmetry and its associated degeneracies are approximate
due to relativistic effects whose size are of order $m_e\alpha^4$, where
$\alpha$ is the fine-structure constant and $m_e$ the electron mass.
Is there a relativistic quantum field theory which has an exact symmetry generalizing
the conservation of the Laplace-Runge-Lenz vector?
In this Letter we will show that such a system exists and use the additional symmetry to facilitate the calculation of its spectrum.

To understand how to formulate the symmetry (\ref{Laplace-Runge-Lenz}) relativistically, it will be
helpful to recall the classic work by Wick and Cutkosky~\cite{WickCutkosky}.   
These authors studied the relativistic Bethe-Salpeter equation for a bound-state wave function $\psi$,
\be
 \psi(p) =  \int \!   \frac{ -4i\lambda \,m_1m_3\,\psi(q) \, {d^4q}/{(2\pi)^4}}{(p{-}q)^2 \big[ (q{-}y_1)^2+m_1^2\big]\big[(y_3{-}q)^2+m_3^2\big]}\,,
 \label{Bethe-Salpeter}
\ee
where $(y_3{-}y_1)^\mu= P^\mu$ is the total four-momentum of the bound state and $(q{-}y_1)^\mu$ and $(y_3{-}q)^\mu$
are the momenta of its two constituents.
This is a natural relativistic generalization of the Schr\"odinger equation
and arises as the approximation to
electron-proton scattering where one retains only all planar ladder diagrams and treats the photon as a spin-0 particle.

Wick and Cutkosky noticed that the equation is invariant under a larger symmetry than the expected $SO(3)$ rotations.
In modern language, their findings may be summarized by the statement that
Eq.~(\ref{Bethe-Salpeter})
is covariant
under the transformations
\be\label{transformations1}
\begin{aligned}
\hspace{-0.25cm} \delta_\xi p^\mu &= 2(\xi{\cdot}p) p^\mu - p^2\xi^\mu\,, \;\;  \delta \psi(p) = -2(\xi{\cdot}p)\, \psi(p)\,,
\\
\hspace{-0.25cm}\delta_{\xi} y_i^\mu &= 2(\xi{\cdot}y_i) y_i^\mu - \big(y_i^2 +m_i^2\big)\xi^\mu\,, \;\;
 \delta m_i = 2(\xi{\cdot}y_i) m_i\,.
\end{aligned}
\ee
These transformations have a simple interpretation as conformal transformations of the {\it momentum} space of the theory.
Following recent literature, we will refer to them as dual conformal transformations. 
Noticing that Eq.~(\ref{Bethe-Salpeter}) is also invariant under translations of $(p,y_i)$
as well as under Lorentz transformations,
one may see that the equation is covariant under a full $SO(4,2)$ group.

The transformations (\ref{transformations1})  can be used to relate solutions which correspond
to different masses. In fact, they imply that the dynamics
depends only on the combination \cite{WickCutkosky}
\be
 u = \frac{4m_1 m_3}{-s + (m_1-m_3)^2}\,. \label{cross-ratio_u}
\ee
This provides a generalization of the concept of reduced mass to this 
particular relativistic setup.
The remaining nontrivial predictions of the $SO(4,2)$ symmetry arise from the subgroup which preserves the masses and $y_i$.

This subgroup is six dimensional since $SO(4,2)$ is 15 dimensional and ten constraints are imposed, but only nine
are independent.
Explicitly, in a rest frame where $y_1=0$ and $y_3=(P^0,\vec{0})$,
and setting $m_1=m_3=m=2\mu$ without loss of generality, we find that the nontrivial generators 
are the following combinations of (\ref{transformations1}), Lorentz boosts, and translations:
\be
\begin{aligned}
\delta'_{\vec \xi} \vec{q} &= \vec{\xi}\left( q_0^2-\vec{q}\,{}^2-P^0q^0+m^2\right)+2\vec{\xi}{\cdot}\vec{q}\,\vec{q}\\
\delta'_{\vec \xi} q^0 &= \vec{\xi}{\cdot}\vec{q} \left( 2q^0-P^0\right)\,,
\end{aligned}
\label{O(4)transformation}
\ee
with a similar transformation for $p$. By construction, the points ($y_i,m_i$) 
are invariant under this symmetry.

These transformations can be interpreted more easily by taking the nonrelativistic limit of the model.
It is well known that Eq.~(\ref{Bethe-Salpeter}) reduces to the Schr\"odinger equation in this limit,
a fact which can be demonstrated by approximating the frequency integration by its residue on the
$1/(q^2+m^2)$ propagator.
Substituting the value of $q^0$ on the residue, $q^0\approx m + {\vec{q}\,{}^2}/({2m})$, 
the transformation (\ref{O(4)transformation}) is reduced to
\be
 \delta'_{\vec{\xi}} \vec q = -\vec{\xi}\left[m (P^0-2m)+\vec{q}\,{}^2\right] +2\vec{\xi}{\cdot}\vec{q}\,\vec{q}\,. \label{NR_O(4)_transformation}
\ee
It is easy to see that this is the canonical transformation generated by the
Laplace-Runge-Lenz vector (\ref{Laplace-Runge-Lenz}) \footnote{
The canonical transformation generated by $2\vec{\xi}{\cdot}\vec{A}$ reproduces Eq.~(\ref{NR_O(4)_transformation})
up to a time shift $\delta_{\vec{\xi}} t=-\vec{\xi}{\cdot}\vec{x}$,
consistent with the fact that we 
considered only the time-independent Bethe-Salpeter equation (\ref{Bethe-Salpeter}).
}.
This demonstrates that the symmetries (\ref{O(4)transformation}),
which arose from $SO(4,2)$ conformal transformations
in momentum space,
are nothing but a relativistic generalization
of the Laplace-Runge-Lenz vector.
For more details on the interpretation of the latter, we refer to \cite{Goldstein}.

Unfortunately, the Wick-Cutkosky model does not define a consistent relativistic theory, as the ladder approximation is not unitary
and lacks multiparticle effects.
Remarkably, a consistent quantum field theory generalizing the above symmetry does exist.
It has been observed that maximally supersymmetric Yang-Mills theory
(${\mathcal N}=4$ SYM) with gauge group $SU(N_{c})$, 
which has a superconformal symmetry, also has a dual momentum space
version of this symmetry, in the planar limit~\cite{DCS}.
(The planar limit, which we are going to work in, is defined by $N_{c} \to \infty$, with the `t Hooft coupling
$\lambda = g^2 N_{c}$ held fixed.) 
As far as we are aware, this is the only known example of a four-dimensional quantum field theory with such a symmetry. 

\begin{figure}[t]
\includegraphics[width=70mm]{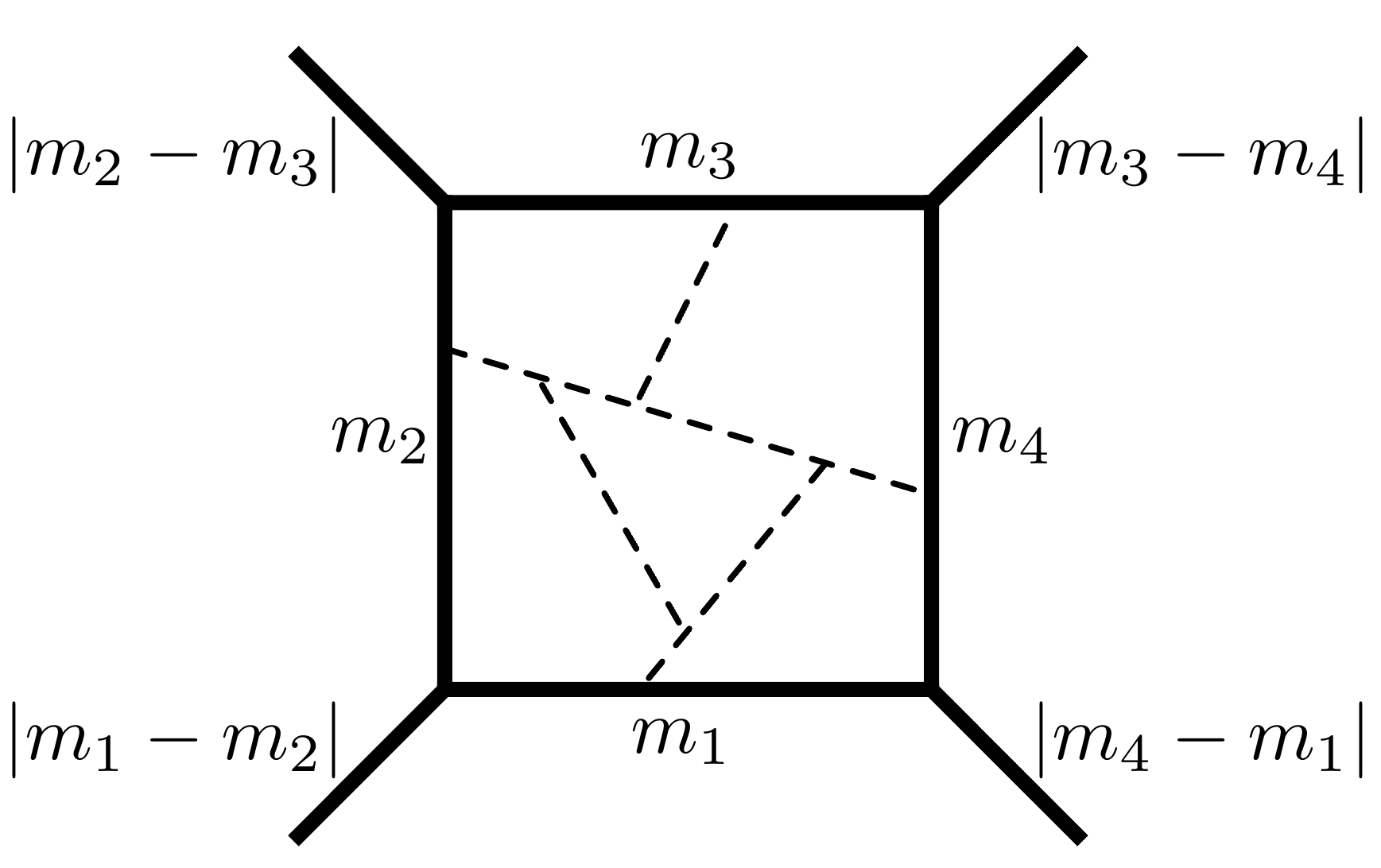}%
\caption{Four-point amplitude in $\mathcal{N}=4$ SYM with nontrivial scalar vacuum expectation values. 
Thick lines correspond to massive $W$ bosons, while dashed lines correspond to massless particles.
\label{fig:amplitude}}
\end{figure}

In the usual formulation, this is a theory of massless particles. However, massive particles can be introduced in a natural way 
via the Higgs mechanism.
This allows us to  discuss the scattering of massive $W$ bosons.
Their masses can be freely adjusted by varying scalar field expectation values.
Let us focus on the four-particle scattering amplitude
depicted in Fig.~\ref{fig:amplitude}.
This amplitude is finite in the ultraviolet, due to the finiteness of $\mathcal{N}=4$ SYM, as well as in the infrared, thanks to the particle masses.
Dual conformal symmetry implies that the dependence on the kinematical invariants and masses is as follows,
for the symmetry breaking pattern $SU(N_{c})\to SU(N_{c}-4)\times U(1)^4$ \cite{Alday:2009zm}:
\be
 A_4(s,t,m_1,m_2,m_3,m_4) = A^{\rm tree}_4 \times M(u,v)\,, \label{covariance_of_M}
\ee
where, as a generalization of Eq.~(\ref{cross-ratio_u}),
\be
 u= \frac{4m_1m_3}{-s+(m_1-m_3)^2}\,,\quad
 v= \frac{4m_2m_4}{-t+(m_2-m_4)^2}\,. \label{cross-ratios}
\ee
In the remainder of this Letter we wish to discuss implications of the structure (\ref{covariance_of_M})
which, as we have seen, is intimately tied to the Laplace-Runge-Lenz vector, regarding the spectrum of the theory.

As depicted in Fig.~\ref{fig:amplitude}, the $W$ bosons interact by exchanging massless gauge fields from the
unbroken part of the gauge group. One can readily see that the interaction is attractive, so they will form bound states.
At weak coupling these are similar to hydrogen states.
As in the Wick-Cutkosky model, we may use Eq.~(\ref{cross-ratios}) to restrict to the case $m_1=m_3=m$.

The exact dual conformal symmetry ensures that the spectrum organizes into complete $SO(4)$ multiplets,
nonperturbatively at any coupling $\lambda$.
The total degeneracy at principal quantum number $n$, including supersymmetry, is $256n^2$.
To extract the spectrum from the amplitude, we will
benefit from relativity by making use of Regge theory \cite{Regge}.
The latter instructs us to group the highest-spin state at each
energy $E_n$ into a trajectory $j(s)$, where $j$ is the spin:
\be
  j(s_n)+1 = n\quad\mbox{when} \quad s_n=E_n^2\qquad\mbox{($n=1,2,\ldots$)}\,. \label{spectrum}
\ee
The analytic continuation of the function $j(s)$ then determines
the behavior of the amplitude in the ultrarelativistic 
limit $t\to\infty$ with $s<0$ fixed,
through $M \sim t^{j(s) +1}$ [provided only that $j(s)$ remains the leading trajectory in that region].
Conversely, if one knows $j(s)$ by some means,
Eq.~(\ref{spectrum}) can be used to determine the spectrum.

\begin{figure}[t]
\includegraphics[width=80mm]{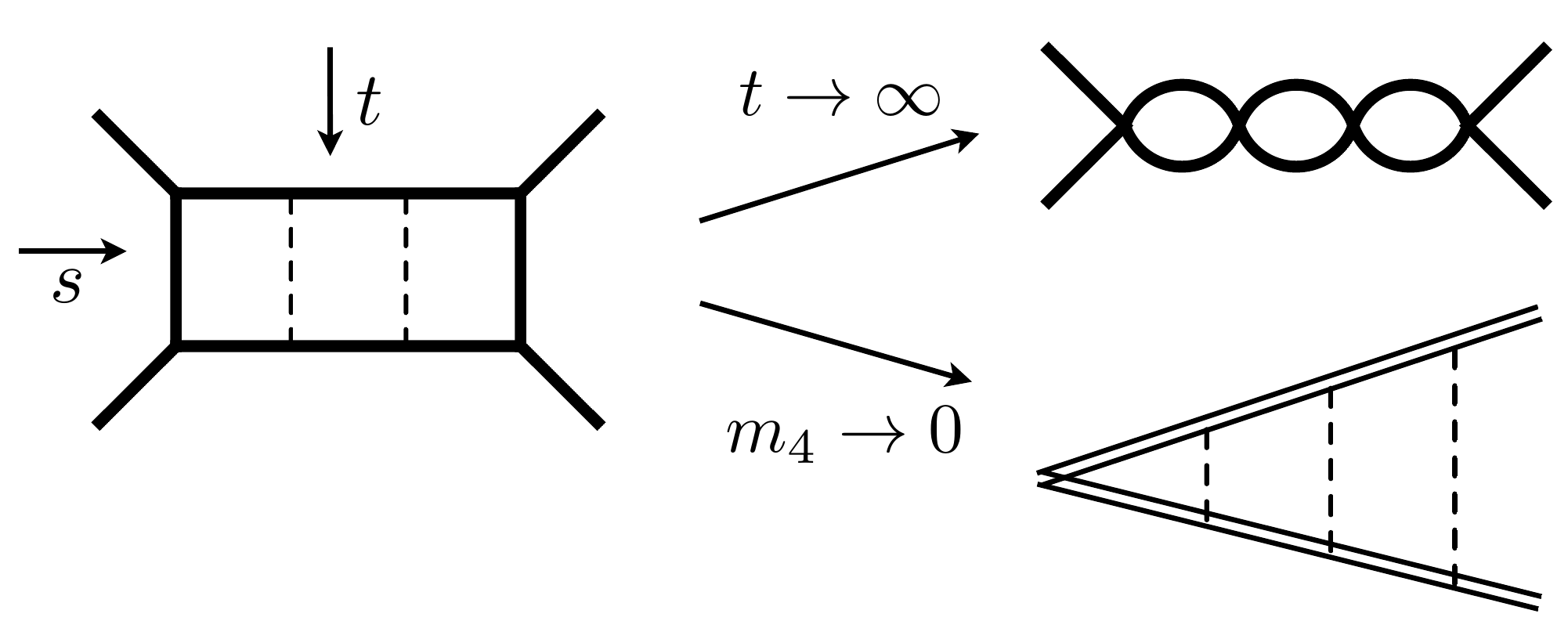}%
\caption{
Different limits of the four-point amplitude that are equivalent thanks to dual conformal symmetry.
The double lines denote Wilson lines.
\label{fig:limit}}
\end{figure}

A traditional way to calculate Regge trajectories perturbatively is to sum
logarithmically enhanced graphs. For example, at the leading-logarithmic accuracy,
the ladder integrals shown in Fig.~\ref{fig:limit} dominate and exponentiate in a simple way.
The exponent, the 
gluon Regge trajectory $j(s)$, is given by a two-dimensional bubble integral. In principle this calculation could be carried out to subleading orders as well (see, e.g., \cite{Henn:2010bk,Correa:2012nk}).

The relativistic Laplace-Runge-Lenz symmetry offers a novel and easier way to
calculate the Regge trajectory $j(s)$.
Through Eq.~(\ref{cross-ratios}),
we see that the limit $t\to\infty$ of the amplitude, with all other variables held fixed, is {\it equivalent} to the limit $m_4\to 0$.
In this limit the amplitude is known to become infrared divergent and its leading terms are 
governed by the anomalous dimension $\Gamma_{\rm cusp}$ of a Wilson loop with a cusp \cite{Korchemsky:1991zp}, $M\sim (m_4)^{\Gamma_{\rm cusp}(\phi)}$.
Equating the exponents in the two asymptotic limits using Eq.~(\ref{cross-ratios}), we thus find that
\be\label{cusp_regge_duality}
 j(s)+1 = -\Gamma_{\rm cusp}(\phi)\quad\mbox{where}\quad s= 4m^2\sin^2\frac{\phi}{2}\,.
\ee
This relation has been derived and used previously in Refs.~\cite{Henn:2010bk,Correa:2012nk}, to which we refer the reader for more details. A similar relation is known to give the infrared-divergent part of the gluon trajectory
as $m^2\to 0$ \cite{Korchemskaya:1996je},
but we
stress that in planar $\mathcal{N}=4$ SYM, Eq.~(\ref{cusp_regge_duality}) holds for the complete function of $s/m^2$.

We wish to combine this relation with
Eq.~(\ref{spectrum}) as a means to obtain the spectrum of hydrogenlike bound states in this theory.
At the lowest order the cusp anomalous dimension is readily calculated by a one-loop graph that corresponds to one rung
in the Wilson line integral of Fig.~\ref{fig:limit} 
\footnote{The formula differs from the corresponding QCD one
\cite{Polyakov:1980ca}
due to the coupling of the Wilson lines to scalars.
},
\be
 \Gamma_{\rm cusp}(\phi) = - \frac{\lambda}{8\pi^2}\, \phi \, \tan  \frac{\phi}{2} + \cO(\lambda^2) \,.
\label{one_loop_cusp}
\ee
To obtain the spectrum we need to solve Eq.~(\ref{spectrum}) or, equivalently, $\Gamma_{\rm cusp}(\phi_n) = -n$.
From Eq.~(\ref{one_loop_cusp}) we see that, since $\lambda$ is small, the solution can occur only for $\phi$ close to $\pi$.  In this region we have
\be
 \Gamma_{\rm cusp}(\pi-\delta) \approx -\frac{\lambda}{4\pi \delta} \,,\label{one_loop_approx_form}
\ee
so that $\delta_n\approx {\lambda}/({4\pi n})$.
Converting to an energy using Eqs.~(\ref{spectrum}) and (\ref{cusp_regge_duality}), we thus find 
\be\label{leadingspectrum}
E_n-2m=-\frac{\lambda^2m}{64\pi^2n^2}+O(\lambda^3)\,.
\ee
This is the well-known hydrogenlike spectrum associated with Eq.~(\ref{Hamiltonian}), as expected, giving a first confirmation of the method.

Because the present hydrogenlike system is embedded in a relativistic quantum field theory
we expect the spectrum to be sensitive to a rich set of multiparticle effects.  For example,
one expects large logarithms to arise from so-called ultrasoft virtual particles,
in an analogy with the computation of the Lamb shift in QED.
These are modes which are infrared compared to the atomic radius, but not  compared to the binding energies.
In fact, as we will see, closely related effects do appear in the computation of $\Gamma_{\rm cusp}$ at the next order,
which makes a nontrivial resummation necessary.

To carry out this resummation systematically, we borrow methods used in the study of the heavy quark static potential in QCD~\cite{Pineda:2007kz}. 
But first we will need to use conformal symmetry one more time,
now in the coordinate space of the theory. Through radial quantization,
conformal symmetry equates the anomalous dimension $\Gamma_{\rm cusp}(\phi)$
to the energy of a pair of static heavy quarks on $\mathbb{S}^3 \times \mathbb{R}$, where the ``time'' $r\in \mathbb{R}$
is the radial distance from the cusp and $\delta$ is the distance between the two quarks on the
 sphere~\cite{Drukker:2011za,Correa:2012nk}.
Combined with the duality (\ref{cusp_regge_duality}), we thus have a relation between {\it dynamical} quarks in flat space,
and {\it static} quarks in the curved space $\mathbb{S}^3 \times \mathbb{R}$.
Such relations (in flat space) are generic
in the large mass limit, but we wish to stress that here we are not taking such a limit and we are discussing the full, relativistic system.
This mapping to the cylinder $\mathbb{S}^3 \times \mathbb{R}$ helps apply standard methods because one is now computing
a static potential.

In the regime $\delta\sim \lambda$ relevant to the bound states,
there are two important length scales on the cylinder: the small size of the pair
and the (unit) radius of the sphere, the latter being comparable to 
the singlet-adjoint energy splitting ${\lambda}/({4\pi\delta})$.
This second fact signals the need for a resummation of perturbation theory.
This was carried out to the next-to-leading order in Ref.~\cite{Correa:2012nk}, whose results we borrow:
\begin{equation}
\begin{aligned}
\Gamma_{\rm cusp}(\pi-\delta) =& \frac{-\lambda}{4\pi\delta}\left(1-\frac{\delta}{\pi}\right) + \frac{\lambda^2}{8\pi^3\delta} \log \frac{\epsilon_{\rm uv}}{2\delta} \\
& \hspace{-1.8 cm}- \frac{\lambda}{4\pi^2}\int_{\epsilon_{\rm uv}}^{\infty} \frac{d\tau}{\cosh(\tau)-1}\left(e^{-\tau\frac{\lambda}{4\pi\delta}}-1\right) + \cO(\lambda^3)\,.
\end{aligned} \label{resummation}
\end{equation}
Here $\epsilon_{\rm uv}$ is a small ultraviolet regulator, which cancels against a divergence of the integral.
In fact we were able to perform the latter analytically.
Upon equating the left-hand side to minus an integer we obtain
the following correction to Eq.~(\ref{leadingspectrum}):
\begin{align} \label{nlospectrum}
\!\!\!\!\left(E_n{-}2m\right)|_{\lambda^3} \!=&
\frac{-\lambda^3m}{64\pi^4 n^2} \left[ S_1(n)+\log\frac{\lambda}{2\pi n}-1-\frac{1}{2n}\right]\!,\!\!
\end{align}
for $n=1,2,3,\ldots$, and where $S_1(n)=\sum_{k=1}^n\frac{1}{k}$ is the harmonic number.

Let us discuss this equation.
First, we note that the size of the correction is uniformly bounded as a function of $n$, and therefore for small $\lambda$
it is always smaller than the leading term given in Eq. (\ref{leadingspectrum}).
This demonstrates that the
perturbative expansion is under control.

Second, we notice the nonanalytic dependence on the coupling through the $\log\lambda$ term.
This originates from the ultrasoft modes alluded to earlier and is conceptually similar to the $(m_e\alpha^5\log\alpha)$
contribution to the Lamb shift in QED. 
It appears earlier by two powers of the coupling in the present model
because ultrasoft scalar exchanges are not dipole suppressed.

Third, the square bracket 
becomes constant at large $n$. Its value is in perfect agreement
with replacing the coupling $\lambda$ in Eq.~(\ref{leadingspectrum})
with the (flat space) static potential,
$\lambda\mapsto \lambda+ \frac{\lambda^2}{2 \pi^2}\left(\log \frac{\lambda}{2\pi}+\gamma_{\rm E} -1 \right)+O(\lambda^3)$
\cite{Pineda:2007kz}, as it should be.

Finally, we wish to mention that
we have verified Eq.~(\ref{nlospectrum}) against a direct next-to-leading order calculation of the spectrum using conventional methods \cite{ToAppear}.
This confirms, in a nontrivial way, that the method based on Eqs.~(\ref{spectrum}) and (\ref{cusp_regge_duality}) provides the correct spectrum.

The duality (\ref{cusp_regge_duality}) can also be verified at strong coupling.
The cusp anomalous dimension
was obtained in semianalytic form in Ref.~\cite{Drukker:1999zq} while
the spectrum was obtained in Ref.~\cite{Kruczenski:2003be}
by solving numerically a differential equation~\footnote{
Reference \cite{Kruczenski:2003be} considered $\mathcal{N}=4$ SYM coupled to a
$\mathcal{N}=2$ matter multiplet, but their specific string solutions apply to our setup as well.},
both using  the AdS/CFT correspondence.
The two formulations appear very different and we were not able to find an analytic match between them.
Nonetheless, when we evaluated numerically the two formulas throughout the range $0<E<2m$
(corresponding to $0<\phi<\pi$), we found perfect agreement within numerical accuracy. 

\begin{figure}[t]
\includegraphics[width=90mm]{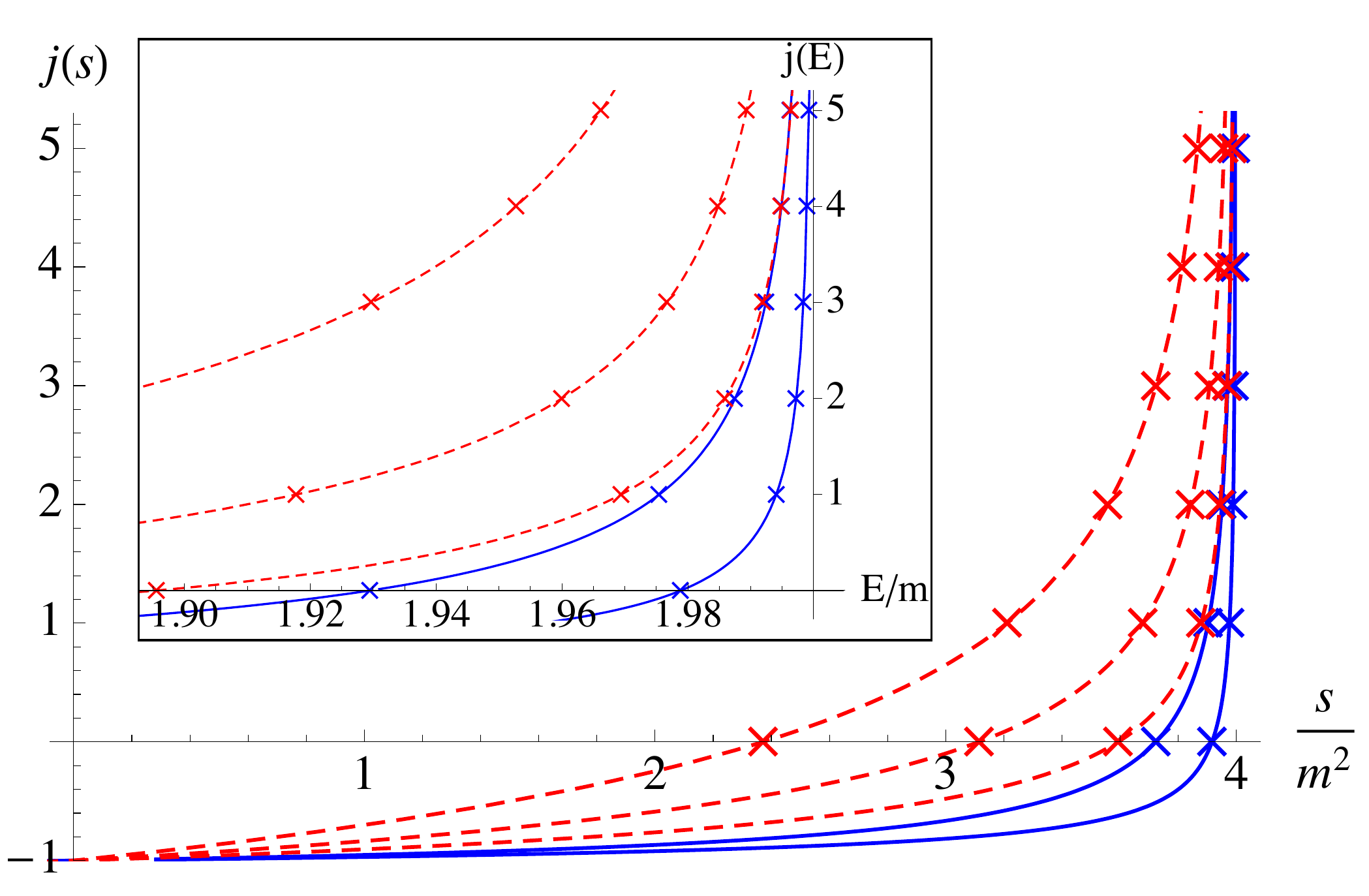}%
\caption{
Regge trajectories of Hydrogen-like states in $\mathcal{N}=4$ SYM theory
for $\lambda=5,10,10,30,100$ (bottom to top).  The solid-blue lines use the weak-coupling formulas
while the dashed-red lines use the large-$\lambda$ formulas (see text).
The bound states (crosses) are obtained by equating $j$ to an integer.
The inset shows the same curves with the total energy in units of mass on the horizontal axis.
\label{fig:ChewFrautschi}}
\end{figure}

In Fig.~\ref{fig:ChewFrautschi} we show the next-to-leading order trajectory
at weak coupling \footnote{To be precise, the plot shows the two-loop fixed-order result of Ref.~\cite{Correa:2012nk},
plus the difference between the resummed expression (\ref{resummation}) and the fixed-order result
expressed only as a function of the one-loop $\Gamma_{\rm cusp}$ in Eq.~(\ref{one_loop_cusp}),
so as to make it well behaved for all values of the angle.},
as well as the strong coupling formula
taken from either one of Refs.~\cite{Drukker:1999zq,Kruczenski:2003be}.
The spectrum is obtained from the curves by solving $j_n(s_n)=n{-}1$ [see Eq.~(\ref{spectrum})].
With increasing coupling the ground state becomes more tightly bound, as expected.
The reader should not attribute a deep meaning to the agreement of the two curves
at $\lambda=10$ and large spin;
this is simply due to the fact that the weak and strong coupling extrapolations of
the flat space static potential turn out to cross roughly at this value.
The difference in shape between
the two curves offers one measure of the current uncertainties at intermediate coupling.
At small $s$ 
the slope is known exactly \cite{Correa:2012at}.

As a final application, the Laplace-Runge-Lenz
symmetry allows us to extend the conventional $SO(3)$ partial wave decomposition
of the four-particle amplitude so as to account for the contribution of full $SO(4)$ multiplets, reducing the complexity of the expansion.
By analyzing the three-loop results from Ref.~\cite{Caron-Huot:2014lda} in this way, we found evidence
that the first subleading power correction in the high-energy limit is controlled by a single Regge pole
or, equivalently, via Eq.~(\ref{cusp_regge_duality}), a single operator of dimension
\begin{align}
\Gamma_1(\phi) =& 
 1 + {\lambda}/({4 \pi^2}) + \mathcal{O}(\lambda^2)\,.
\end{align}
Details of the analysis and the full three-loop trajectory will be reported elsewhere \cite{ToAppear}.
This simplicity hints at further structure in the dynamics of this model,
which does not directly follow from the Laplace-Runge-Lenz symmetry but which the latter may help uncover.

To conclude, we mention that the cusp anomalous dimension in $\mathcal{N}=4$ SYM has recently been
reformulated in terms of a system of integral equations which embody the integrability of this theory~\cite{Correa:2012hh}.
Combined with the present results, this could lead to an exact determination of the spectrum at finite coupling
in this interacting quantum field theory.

{\it Acknowledgments}
J.~M.~H. is supported in part by DOE Grant No.~DE{-}SC0009988.
S.~C.~H.'s work was supported in part by NSF Grant No.~PHY-1314311. 
Both authors acknowledge support from the Marvin L. Goldberger Fund.

\end{document}